\documentclass{elsart}

\usepackage{amssymb,graphicx}

\journal{Physics Letters B}

\def\Journal#1#2#3#4{#1 #2 (#3) #4}

\def\PLB{{Phys. Lett.}  B}
\def\PR{Phys. Reports}

\def\PRD{{Phys. Rev.} D}

\def\EPC{{Eur. Phys. Journal} C}

\def\lsim{\mathrel{\rlap{\lower4pt\hbox{\hskip1pt$\sim$}}
    \raise2pt\hbox{$<$}}} %less than or approx. symbol
\def\gsim{\mathrel{\rlap{\lower4pt\hbox{\hskip1pt$\sim$}}
    \raise2pt\hbox{$>$}}} %greater than or approx. symbol
\def\gg{$\gamma\gamma$}
\def\beq{\begin{equation}}
\def\enq{\end{equation}}

\begin{document}

\begin{frontmatter}

% Title, authors and addresses

% use the thanksref command within \title, \author or \address for footnotes:
% \title{Title\thanksref{label1}}
% \thanks[label1]{}
% \author{Name\thanksref{label2}}
% \thanks[label2]{}
% \address{Address\thanksref{label3}}
% \thanks[label3]{}
% including your email address:
% \address{Address\thanksref{email}}
% \thanks[email]{E-mail: }

\title{Tagging Two-Photon Production at the LHC}

% use optional labels to link authors explicitly to addresses:
% \author[label1,label2]{}
% \address[label1]{}
% \address[label2]{}

\author{K. Piotrzkowski\thanksref{email}}

\thanks[email]{E-mail: krzysztof.piotrzkowski@cern.ch}
\thanks[new]{Now at Universit\'e catholique de Louvain, FYNU Lab., 
B--1348 Louvain-la-Neuve}
\address{CERN, EP Division, CH--1211 Geneva 23, Switzerland \\and\\
Institute of Nuclear Physics, Kawiory 26A, PL--30055 Krak\'ow, Poland
\thanksref{new}}

\begin{abstract}
Tagging two-photon production offers a significant 
extension of the LHC physics programme. Effective luminosity of high-energy
\gg\ collisions reaches 1\% of the proton-proton luminosity and 
the standard detector techniques used for measuring very forward proton
scattering should allow for a reliable extraction of interesting 
two-photon interactions. Particularly exciting is a possibility of
detecting two-photon exclusive Higgs boson production at the LHC.
\end{abstract}

%\begin{keyword}
% keywords here, in the form: keyword \sep keyword

% PACS codes here, in the form: \PACS code \sep code
%\PACS 
%\end{keyword}
\end{frontmatter}

% main text
\section{Introduction}
\label{int}
Two-photon physics has been traditionally studied at
$e^+e^-$ colliders owing to large fluxes of virtual photons
associated with the beams. However, at the LHC for the first
time the proton beam energy will be so high that the effective 
luminosity of \gg\ collisions will permit performing
meaningful and important experiments. The measurements of 
high-energy \gg\ collisions at the LHC are intrinsically very 
interesting, and to large extend are complementary to the `base-line' 
$pp$ studies.

In this paper, an experimental feasibility of tagging two-photon 
production in proton-proton collisions at the LHC is
considered. The effective \gg\ luminosity of the tagged
two-photon collisions is evaluated and used to estimate the physics 
potential for such measurements.
\section{LHC as a \gg\ collider \label{gg}}
For majority of two-photon processes the equivalent photon 
(or Weizs\"acker-Williams) approximation (EPA) can be
successfully applied \cite{budnev}. In EPA two-photon production at 
the LHC proceeds in two steps: first, two photons are emitted by
incoming protons and then, the photons collide producing a 
system $X$ while the protons remain intact in the $elastic$ production,
$pp\rightarrow (\gamma\gamma\rightarrow X)\rightarrow ppX$, or
one of them dissociates into a state $N$ in the $inelastic$ production,
$pp\rightarrow (\gamma\gamma\rightarrow X)\rightarrow pNX$ 
\footnote{Third class of events when two protons dissociate is
not considered here.}.
%see Fig. \ref{fig:graph}. 
Hence, the proton-proton cross-section 
is a product of the photon-photon cross-section and two 
photon spectra:
$$
{\rm d}\sigma_{pp}=\sigma_{\gamma\gamma}~{\rm d}N_1~{\rm d}N_2\ .
$$

In EPA the photon spectrum is a function of the photon energy 
$\omega$ and its virtuality $Q^2$  \cite{budnev}:
\beq
\label{eq:2}
{\rm d}N=\frac{\alpha}{\pi}\frac{{\rm d}\omega}{\omega}\frac{{\rm d}Q^2}{Q^2}
\left[\left(1-\frac{\omega}{E}\right)
\left(1-\frac{Q^2_{min}}{Q^2}\right)F_E
+\frac{\omega^2}{2E^2}F_M\right],
\enq
where $\alpha$ is the fine-structure constant, $E$ is the incoming
proton energy and the minimum photon virtuality 
$Q^2_{min}\simeq[M_N^2E/(E-\omega)-M_p^2]\omega/E$, where $M_p$ is 
the proton mass and $M_N$ is the invariant mass of the final state $N$.
For the elastic production, assuming the dipole 
approximation for proton form factors, $F_M=G_M^2$ and 
$F_E=(4M_p^2G_E^2+Q^2G_M^2)/(4M_p^2+Q^2)$,  and 
$G_E^2=G_M^2/7.78=(1+Q^2/0.71{\rm GeV}^2)^{-4}$. For the inelastic 
production $F_M=\int{\rm d}x F_2/x^3$ and $F_E=\int{\rm d}x F_2/x$, 
where $F_2(x,Q^2)$ is the proton structure function and
$x\simeq Q^2/M_N^2$.
The spectrum is
strongly peaked at low $\omega$, therefore the photon-photon center of mass 
energy $W\simeq 2\sqrt{\omega_1\omega_2}$ is usually much smaller 
than the total center of mass energy which equals $2E=14$~TeV. 
For the elastic production the photon virtuality is usually low, 
$\langle Q^2\rangle\approx0.01$~GeV$^2$, therefore
the proton scattering angle is very small, $\lsim20~\mu$rad.  

\begin{figure}[h]
%\begin{center}
\includegraphics*[width=15cm]{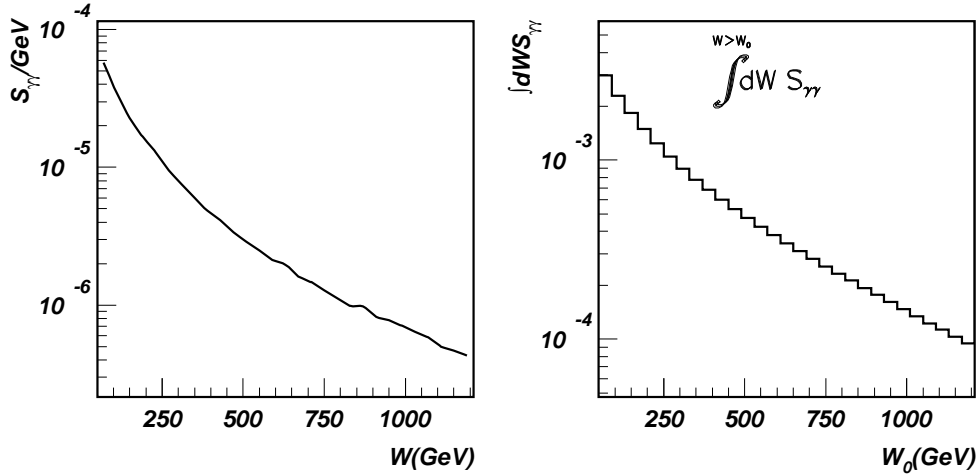}
%\end{center}
\caption{Elastic \gg\ luminosity spectrum and its integral 
$\int^{W>W_0}dW S_{\gamma\gamma}$ at the LHC, for the range of
photon energy and virtuality, 5~GeV~$<\omega<7000$~GeV and 
$Q^2_{min}<Q^2<2$~GeV$^2$.\vspace{0.3cm}}
\label{fig:lumi}
\end{figure}

The luminosity spectrum of photon-photon collisions, $S_{\gamma\gamma}$, 
can be introduced in EPA by integration of the product of the photon
spectra given by Eq. \ref{eq:2} over the photon virtualities and energies 
keeping fixed $W$. As an example, in Fig. \ref{fig:lumi}, assuming for each 
photon an integration interval of 5~GeV$<\omega<E$ and 
$Q^2_{min}<Q^2<2$~GeV$^2$, $S_{\gamma\gamma}$ and its 
integral $\int^{W>W_0}dW S_{\gamma\gamma}$ are shown for the elastic
production as a function of $W$ and the lower integration limit, 
$W_0$, respectively.
The integrated spectrum directly gives a fraction of the $pp$ LHC luminosity
available for the photon-photon collisions at $W>W_0$.
It is remarkable that for $W_0=50$~GeV this fraction is close to 1\%, 
and for the nominal $pp$ luminosity of $10^{34}$~cm$^{-2}$s$^{-1}$ 
the \gg\ luminosity at $W>200$~GeV is $3\times10^{31}$~cm$^{-2}$s$^{-1}$.
%and is similar to what has been so far achieved at Tevatron or HERA. 
For the inelastic two-photon production $S_{\gamma\gamma}$ is even
larger and will be discussed below.
\section{Tagging two-photon production}
\label{tag}
Tagging two-photon production at the LHC would serve two purposes.
First, when both elastically scattered protons are detected 
({\it double tag}) the \gg\ center of mass energy $W$ can be measured
independently and improve an overall reconstruction of the final state $X$.
Secondly, tagging is needed in order to suppress backgrounds and 
to obtain a clean two-photon data sample. In particular, the 
proton scattering angle can be used to extract the two-photon
signal both for the double and {\it single} tags when only one
very forward proton is detected.
 
%\section{Measurement of forward protons at the LHC}
At the nominal running conditions the LHC beam at the interaction
point (IP) has the Gaussian lateral widths 
$\sigma^*_x$=$\sigma^*_y$=$16~\mu$m 
and the angular divergence in horizontal and vertical planes 
$\sigma_{\theta_x}^*$=$\sigma_{\theta_y}^*$=$32~\mu$rad. However, for the 
initial running at medium luminosity of $10^{33}$~cm$^{-2}$s$^{-1}$
almost two times smaller lateral
beam size as well as the angular divergence at the IP are expected. 
At the same time, the event pile-up at the central detectors is not
prohibitively large. The beam energy spread will be $10^{-4}$ \cite{lhc}. 
The beam divergence is comparable with the typical proton scattering angle in 
the two-photon processes hence effectively protons leave the IP 
at zero-angle. These protons have however smaller energy than the beam 
protons and are more strongly deflected in the beam-line magnetic field.
The standard method of measuring such forward scattered protons 
utilizes so-called Roman-pots, position sensitive detectors installed 
far away from the IP in the beam vacuum to allow the closest 
possible approach to a beam. 

The detector layouts for the Roman-pot detectors so far considered by the 
{\sc Totem} \cite{totem} and {\sc Atlas} \cite{atlas} collaborations, mainly
in the context of the total and elastic $pp$ cross-section measurements 
at the LHC, result in a significant acceptance for zero-angle protons 
which lost at least a few \% of their initial energy. It would correspond 
to the tagged photon energies of several hundred GeV and would therefore 
limit the studies of two-photon production only to a domain of very large $W$. 
However, to improve sensitivity to very low angle elastic $pp$ scattering,
it was recently proposed \cite{angeles} to add new detector stations more 
far away ($\approx 240$~m) from the IP. This is also an excellent place for 
tagging two-photon production since at this location, for the nominal LHC 
beam optics in the horizontal plane, the so-called betatron phase advance is 
$\approx\pi$, the beam size has a minimum, and the dispersion $D$ is large, 
about 100~mm. It means that in the horizontal plane a 
detector can approach the beam very closely and at the same time the 
average horizontal displacement $\Delta x$ with respect to the 
beam axis due to the proton energy loss is large, $\Delta x=D\omega/E$.
Therefore, measurement of  $\Delta x$ gives directly 
the tagged photon energy. Additionally, the angle between the 
proton momentum and the beam axis, $\theta_x$, is proportional to the same 
angle at the IP, $\theta_x\simeq\theta_x^*/3$. On the other hand,
in the vertical plane the betatron phase advance is $\approx\pi/2$ at
this location, 
therefore the angle $\theta_y$ is proportional to the proton vertical 
displacement at the IP and the vertical proton displacement 
$\Delta y [{\rm mm}]\simeq0.01\times\theta^*_y [\mu$rad]. 
Hence, the measurement of 
$\theta_x$ and $\Delta y$ directly provide information on proton 
transverse momentum at the IP, $p_T$, hence on photon virtuality 
$Q^2\simeq p_T^2\simeq E^2({\theta_x^*}^2+{\theta_y^*}^2)$.

\begin{figure}[h]
\includegraphics*[width=15cm]{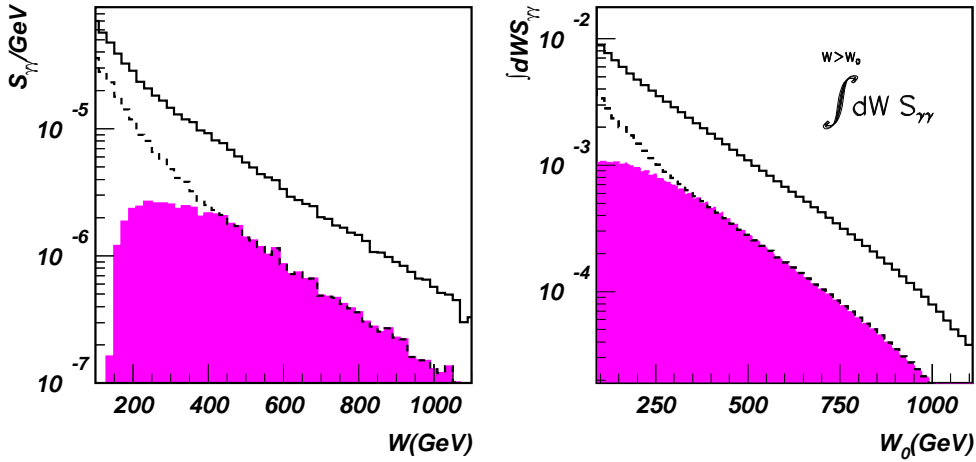}
\caption{Tagged photon-photon luminosity spectrum $S_{\gamma\gamma}$
and its integral $\int^{W>W_0}dW S_{\gamma\gamma}$ assuming double 
tags (shaded histograms) and single tags, for all (solid line) and 
only for elastic (dashed line) events; the tagging range is 
$70<\omega<700$~GeV and  $Q^2_{min}<Q^2<2$~GeV$^2$.\vspace{0.3cm}}
\label{fig:lumi2}
\end{figure}

To ensure enough room for the beam steering and to keep the detectors
in a `shadow' of beam collimators a 1~mm minimum distance between the
detector edge and the beam axis should be assumed \cite{totem}. Since
the proton beam size in the horizontal plane is small,  
$\sigma_x\simeq3\sigma^*_x\lsim50~\mu$m, this distance corresponds to 
more than 20 beam widths, far away from the beam core.
The minimum approach of 1~mm corresponds to the minimum tagged photon energy 
of 70~GeV. The maximum tagged energy is about 700~GeV due to a beam-line
geometrical acceptance, and because for such a large energy loss $D$ starts 
to change with the scattered proton energy, making the
energy measurement less reliable. Assuming this tagged energy range, the 
photon-photon luminosity spectrum is shown in Fig. \ref{fig:lumi2}, 
for the double tagging, when both scattered protons are detected as well
as for the single tagging, when at least one scattered proton is detected;
photon virtuality is restricted to $Q^2_{min}<Q^2<2$~GeV$^2$. It can be noted
that the \gg\ luminosity spectrum for the double tagging is sizable in the 
range $200\lsim W\lsim500$~GeV, whereas the single tagging preserves a major
fraction of the total elastic \gg\ luminosity. Including the inelastic
contribution to the single tagged spectrum increases $S_{\gamma\gamma}$
approximately by a factor of three, assuming the maximum dissociative mass 
$M_N$ of 20~GeV \footnote{This ensures that debris of the system $N$ are
not observed in the central detectors.}.
 
The required detector size is small, with a sensitive area of each detector
plane of the order of 2~cm$^2$, with excellent spatial resolution
in the 10--20~$\mu$m range. There should be two detector stations 
separated by 2--4~m to ensure a precise measurement of direction of
the proton momentum. Modern silicon micro-strip or pixel sensors are the most 
probable candidates. The crucial alignment of the detector sensors 
with respect to the beam axis should be possible using the elastic $pp$ 
events when protons cross several detector planes. Setting the final
momentum scale requires a precise knowledge of the integral of
the magnetic field along the scattered proton trajectory, however the
photon energy scale might be set using the data where the final state $X$
is fully detected in the central detectors and a precise and independent
determination of $W$ is possible.

\begin{figure}[b]
\includegraphics*[width=15cm]{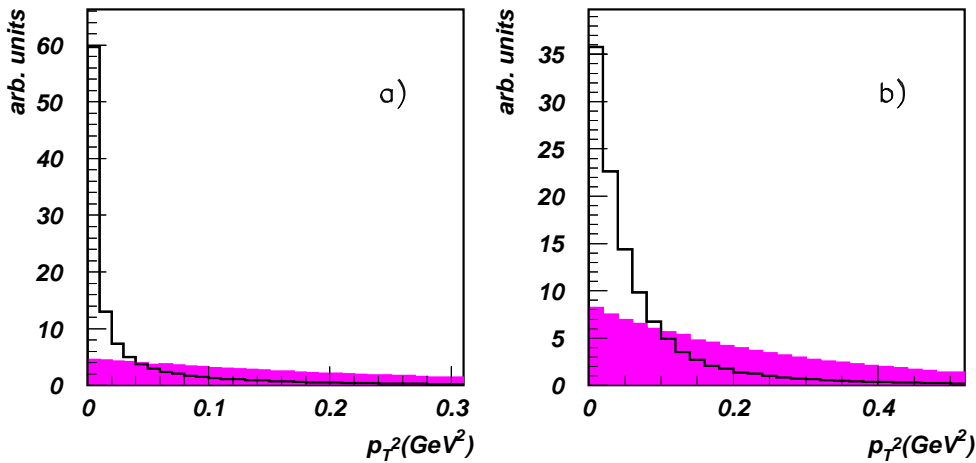}
\caption{a) Distribution of the transverse momenta squared of the scattered 
protons for the two-photon (empty histogram) and pomeron-pomeron 
(shaded histogram) collisions assuming the diffractive slope 
$b$=4~GeV$^{-2}$; b) the same distributions smeared by the beam divergence
for the initial running conditions. Distributions have the same normalization
for $p_T^2<2$~GeV$^{2}$ and correspond to a 100~GeV proton energy loss.
\vspace{0.3cm}}
\label{fig:smear}
\end{figure}

For the above spatial resolutions, the final resolution on  
$W$ and photon virtualities would be determined by the geometrical beam 
properties at the IP. In this case, the horizontal displacement due 
proton energy loss $\Delta x$ is smeared by $\sigma_x\lsim50~\mu$m leading 
to a 5~GeV uncertainty on $W$=200~GeV, for example.  

The same reaction, $pp\rightarrow ppX$, occurs also in strong 
interactions, via fusion of two colorless objects, $pomerons$, and will
therefore interfere with the two-photon fusion. However,
this so-called central diffraction usually results in much larger transverse 
momenta of the scattered protons, following the distribution
$\exp{(-bp_T^2)}$, with the expected at the LHC diffractive slope 
$b\simeq4$~GeV$^{-2}$~\cite{martin}. Soft pomeron-pomeron interactions have 
several orders of magnitude larger cross-sections than the \gg\ interactions,
but for the hard processes the cross-sections are of similar size (see below).
Therefore, the measurement of proton $p_T$ is vital for extracting the \gg\ 
signal. In Fig. \ref{fig:smear}, the `true' and smeared distributions of 
$p_T^2$ are compared assuming the same cross-sections integrated over
$p_T^2$. It shows that in such a case the two-photon signal is clearly
visible and can be well extracted, and for example, for 
$p_T^2<0.05$~GeV$^{2}$ the pomeron-pomeron contribution 
(neglecting interference effects) is about 20\%.
One should note that for the double tagged events the separation is
even more powerful since one can independently use for that purpose
$p_T$ of each proton.
\section{Examples of tagged two-photon physics potential}
\label{phys}
The exclusive \gg\ production of one or two heavy particles, as for example 
in $\gamma\gamma\rightarrow H,t\bar{t},$ or $W^+W^-$ processes, 
is particularly interesting. These events are clean 
-- two (or one) very forward protons measured far away from the IP and only 
one or two particles produced and decaying in the central detectors,
giving an additional handle for background suppression.

In the exclusive two-photon Higgs boson production in leading order 
the photons  couple via a fermion loop which results in sensitivity 
to any new fermion state, even significantly beyond the $W$ scale, hence to 
possible departures from the Standard Model (SM) physics.

\begin{figure}[h]
%\begin{center}
\includegraphics*[width=15cm]{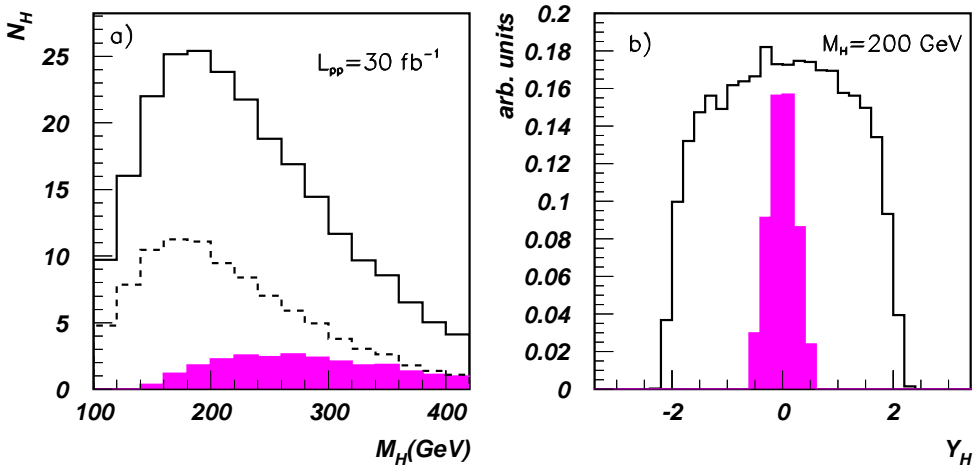}
%\end{center}
\caption{a) Number of the SM Higgs boson events as a function of its mass,
exclusively produced in \gg\ collisions for the integrated $pp$ 
luminosity of 30~fb$^{-1}$, assuming double tags (shaded histograms), 
and single tags for all (solid line) and only for elastic (dashed line) 
scattering; b) Rapidity distribution for the tagged Higgs boson production.
\vspace{0.3cm}}
\label{fig:higgs}
\end{figure}

The number $N_H$ of the two-photon produced Higgs bosons is given by 
\cite{borden}: $$N_H= 
\frac{4\pi^2\Gamma_{\gamma\gamma}}{M_H^2}L_{pp}S_{\gamma\gamma}(W=M_H),
$$
where $L_{pp}$ is the proton-proton integrated luminosity,
$M_H$ is the Higgs boson mass and $\Gamma_{\gamma\gamma}$ is
the  ${H\rightarrow\gamma\gamma}$ width. In Fig. \ref{fig:higgs}, $N_H$ 
expected in SM is shown for the integrated $pp$ luminosity which
corresponds to first three years of the LHC running at medium luminosity. 
It shows that the double-tagged Higgs boson production is statistically
very limited at low $M_H$, but the single-tagged production is
not negligible starting already at $M_H$=100~GeV.

For the same luminosity more than five thousand `gold-plated' double-tagged 
$W^+W^-$ pairs would be exclusively produced at $W>200$~GeV, assuming the
asymptotic value of $\sigma_{\gamma\gamma\rightarrow WW}\simeq200$~pb. 
The corresponding number of the exclusive top pairs 
is unfortunately more than hundred times smaller.
The exclusive $W^+W^-$ production constitutes also an `irreducible'
background for the events when Higgs boson decays into $W^+W^-$. Therefore,
for $M_H\gsim200$~GeV the signal $H\rightarrow ZZ$ seem experimentally
more preferable. Given significant backgrounds and low signal 
statistics, such two-photon measurements cannot possibly be used for a 
Higgs boson search, but they would provide an important handle on 
$\Gamma_{\gamma\gamma}$ at the LHC \cite{elena}.

As was mentioned before, the same event topology have also
the pomeron-pomeron interactions. Recent studies \cite{martin} show 
that the pomeron-pomeron exclusive Higgs boson production has a similar
cross-section to the two-photon case, therefore the two-photon
signal can be statistically extracted using the
distributions of the proton $p_T$. The pomeron-pomeron cross-section
is theoretically not well controlled due to soft final-state 
interactions and the associated {\em survival probability}.
In contrast, provided that $Q^2$ is not too large,
the two-photon cross-section is much less sensitive to these
effects \cite{martin}. On the other hand,
interference between the pomeron-pomeron and \gg\ amplitudes might 
give information (in analogy to Coulomb and elastic $pp$ scattering, 
for example) about the amplitude behaviour at energies beyond available
at the LHC.

Of course, the photon-photon measurements would also extend search for
the new physics at the LHC. In particular, \gg\ search for new
particles expected in supersymmetric models would be
to large extend complementary to the corresponding 
proton-proton studies \cite{zerwas}.

Many interesting QCD studies should also be possible, as for example 
measurements of the exclusive production of multi-jets with large 
transverse energy, 
or the vector meson and photon production at very high transverse 
momenta~\cite{jeff}.
\section{Summary}
The initial studies presented in this paper indicate that the installation 
at about 240~m from the IP of the recently proposed detectors might
permit, at relatively low cost and effort, 
to utilize the LHC as a high-energy \gg-collider.
The significant luminosity of the tagged photon-photon collisions
open in particular an exciting possibility of studying the exclusive
production of the Higgs boson, and for searches of new phenomena in 
high-energy \gg\ collisions. 
\section*{Acknowledgements}
I would like to thank Vincent Lemaitre for very useful discussions
and comments.

\end{document}